**С.Б. Куликов**

# ФИЛОСОФИЯ О ВРЕМЕНИ В ЕСТЕСТВОЗНАНИИ И ГУМАНИТАРНЫХ НАУКАХ С АНАЛИТИЧЕСКИХ ПОЗИЦИЙ

*В статье обсуждаются возможности аналитической философии (идеи Р. Карнапа, в частности) для описания времени в рамках гипотетически возможного единого языка, где отдельные суждения относительно времени раскрывают его целостный смысл, реконструируемый на базе обобщения частных актов осмысления времени в естествознании и гуманитарном знании. Естественнонаучный подход отображен физикой, а гуманитарный – философией.*

## Введение

Основная **проблема** исследования – эксплицировать методологические возможности аналитической философии (как знания об основаниях аргументации) в отношении перспектив интерпретации понятия времени в качестве символа, отображающего пересечения оснований современных философии и физики. В целом исследование выполняется в русле направления, восходящего к идеям Р. Карнапа [1. P. 50–53]. Так, Р. Карнап полагал, что язык как система знаков теоретически может объединить естественные науки и гуманитарное знание, причем в основу исследований закладывалась формальная логика как язык науки в целом. В этом отношении подразумевается некоторый искусственный язык, в рамках которого положения отдельных наук становятся взаимно переводимы. Автор полагает важным выполнить проверку данного допущения, сделав упор на возможности, предоставляемые естественным языком.

Обсуждение заявленной проблематики позволяет достичь **цели** раскрытия оснований, на базе которых формулируются суждения в физике и философии как отдельных областях научного знания, равно как уточнить перспективы построения суждений о времени на тех же основаниях в научном знании в целом. Достижение поставленной цели предполагает решение следующих **задач**:

– выявить в проблематике ключевые направления и сформулировать эти направления в виде базовых тезисов касательно смысла понятия «природа времени»;

– проследить логические следствия базовых тезисов о природе времени и прояснить связи, которые имеются между философией, физикой и (в перспективе) другими научными дисциплинами в плане понимания времени.

В соответствии с представленной последовательностью действий статья включает два основных раздела. Первый раздел посвящен обсуждению вариантов отображения в системе научных понятий и высказываний связи времени в природе с идеей темпоральности человеческого бытия. Во втором разде-ле показаны способы координации времени в природном, социальном и антропологическом его смысле в философии.

## Время в природе и темпоральность человеческого бытия

Предложенный в исследовании подход актуален в отношении вопроса, насколько в принципе возможны представления о времени в виде Универсальных Часов как символе универсального измерительного прибора длительности и изменчивости природных явлений. Целесообразность обсуждения такого вопроса связана с общими следствиями, вытекающими из теории относительности А. Эйнштейна. Релятивистские разработки А. Эйнштейна и его непосредственных предшественников (А. Пуанкаре, Х. Лоренц и др.), согласно которым нет универсальной системы отсчета, а время течет в каждой из систем относительно любой другой в зависимости от близости / удаленности движения этих систем от скорости света, привели к мысли о спорности самой необходимости использования инвариантной временной шкалы. Тем самым под вопросом оказались Универсальные Часы. В философии все это привело к вопросу о возможности универсального стандарта познания как фундамента науки, причем возобладали представления об отсутствии такого фундамента или как минимум его проблематичности. Точки зрения по данному вопросу могут быть сгруппированы следующим образом:

– авторы, описывающие отсутствие универсальных стандартов и оснований [2–4];

– авторы, в том или ином плане признающие необходимость многообразия стандартов и оснований [5–7];

– авторы, ищущие выход из ситуации, когда универсальные стандарты проблематичны, но желательны для измерения изменений в развитии научного познания [8].

Итак, мы можем зафиксировать первый из упомянутых выше базовых тезисов относительно проблемы времени. Тезис носит основополагающий характер, поэтому имеет «нулевую» нумерацию и содержание следующего вида:

*Т.0.* Анализ отображения в научном знании (физико-математическом естествознании) природы времени позволяет прояснить вопрос об основаниях науки в философии и физике в частности и философии науки в целом.

Ход обсуждения в дальнейшем предполагает анализ следующих тезисов:

*Т.1.* Универсальные Часы, понятые как символ, раскрывают путь для понимания общих оснований научного познания в философии науки.

*Т.2.* Установление связи между наукой и человеческим бытием открывает перспективу конституирования философии физики, основанной на темпоральности человеческого бытия.

*Т.3.* Корректность построений в современной философии физики зависит от учета субъективного характера времени, подразумеваемого в работах Э. Гуссерля.

*Т.4.* Символизм Универсальных Часов раскрывает сущность отношений, с одной стороны, между философскими спекуляциями, а также естествознанием и гуманитарным знанием – с другой.

Первые два тезиса обсуждаются в данном разделе. Аргументация третьего и четвертого тезисов представлена во втором разделе статьи. При этом в поддержку *Т.1* автор предлагает следующие доводы:

1. Аргументация, которая имеет силу в рамках современной физики, влечет за собой осознание проблематического характера понятия времени как универсального способа измерения длительностей и изменений и отсылает к следующей аналогии. Время как способ измерения длительности процессов в классической физике отображало вариант реализации общезначимых стандартов. В современной физике наблюдается набор конкурирующих теорий, которые в плане собственных оснований обладают статусом достоверных знаний и в которых время теряет свой универсальный характер. В частности, это общая теория относительности [9] и релятивистская теория гравитации [10]. Проблематично уже само по себе наличие нескольких теорий, равно претендующих на статус истинного описания действительности. Но в отношении проблемы времени такое многообразие теоретических построений фундаментального вида влечет нижеупомянутые последствия.

2. Современная философия науки исследует область действительности, которая непосредственно связана с вопросом о природе научного познания. Этот вопрос подразумевает принципы получения достоверных знаний о материальной природе, мышлении, обществе и других объектах. В частности, это принцип концептуального единства, дающий возможность свести научное знание в целостную систему понятий. И именно здесь актуализируется проблематический характер таких понятий, как пространство и время. В частности, пространство является одним из основных понятий для географии и геометрии. В географии пространство совпадает с местоположением природных объектов, поэтому есть основания определять пространство как «пустоту» [11. P. 139]. В отличие от такого подхода в геометрии пространство совпадает с координацией точек, представляющей собой варианты построения топологического многообразия [12. P. 125–143]. Точно так достаточно проблематично обнаружить универсальные принципы построения единой системы научных понятий касательно времени, очевидно различающегося по смыслу, допустим, в истории, где время означает чередование эпох, и в химии, где время совпадает с мерой скорости протекания химических процессов.

Разнообразие способов понимания времени, которое зависит от вариантов применения этого понятия в различных науках, рожденных человеком при развитии исследовательской деятельности, обусловливает необходимость прояснения предполагаемой связи между понятием времени, наукой и характеристиками человеческого бытия, лежащими в основе выделения единства основ философии физики (и даже философии науки в целом). Это заключение позволяет перейти к обсуждению тезиса, который формулируется следующим образом:

*Т.2.* Установление связи между наукой и человеческим бытием открывает перспективу конституирования философии физики, основанной на темпоральности человеческого бытия.

Автор предлагает такие доводы в поддержку выдвинутого положения:

1. Общая связь науки и человеческого бытия видится в следующем ключе. Результаты, достигнутые в ходе исследований в общей онтологии, могут помочь построить систему научных представлений и в философии науки. Вслед за М. Хайдеггером науку можно истолковать как совокупность утверждений, в целом выражающих темпоральность существования [13. P. 9–10, 371]. Не углубляясь в этот специальный вопрос, поясним, что данное поло-жение, вероятнее всего, следует понимать в качестве указания на стремление к познавательным действиям, призванным разметить, упорядочить конечное бытие, стремящееся к своему завершению.

2. Тезис о конститутивной роли человеческого бытия в отношении философии физики предполагает следующую аргументацию:

Известно, что философы в XX в. раскрывают ключевую роль в понимании человеческой природы понятий языка и времени. Причем к такому результату приходят философы далеких по своим предпосылкам направлений. Сторонник аналитической философии Л. Витгенштейн открывает особый статус языка [14]. Видный представитель феноменологии и экзистенциализма М. Хайдеггер показывает конститутивное значение времени для понимания человеческого существования [13], но и язык как «дом бытия» не обойден вниманием в данном контексте. Несмотря на то, что М. Хайдеггер напрямую не является мыслителем, затрагивающим вопросы философии науки в целом и философии физики в частности, результаты его исследований не лишены эпистемологического значения. В свете этого философия физики должна учитывать особое значение языка и времени для исследовательской деятельности. Причем здесь есть немаловажный нюанс. Время – это одна из основ культурных и исторических процессов, на которой базируются гуманитарные науки. Вместе с тем время – это еще и основа для различения местоположения и порядка естественных процессов, в особенности движения небесных тел. В результате есть два разных понятия для обозначения одного и того же явления, а именно «времени человека» и «времени природы». Но важны здесь не сами по себе отличия между данными понятиями, а то, что значимость проблемы языка становится непреложной, и вопрос о времени оказывается соразмерен коммуникативной сфере, созданной человеком.

Автор заключает, что концептуализация времени и в естествознании, и в гуманитарном знании равно влечет за собой построение образа действительности, соразмерного человеческому бытию.

## Способы координации времени в естественнонаучном, социальном и антропологическом его смысле средствами философии

В этом разделе обсуждаются тезисы *Т.3–Т.4*, которые подразумевают выделение, с одной стороны, связи между субъективистскими (релятивистскими) истолкованиями времени и его естественнонаучного способа понимания, а с другой – раскрытие символизма времени в образе Универсальных Часов и его отношения к основаниям отдельных ветвей научного познания. Первый из обсуждаемых тезисов формулируется следующим образом:

*Т.3.* Корректность построений в современной философии физики зависит от учета субъективного характера времени, подразумеваемого в работах Э. Гуссерля.

Автор предлагает такие аргументы в поддержку выдвинутых положений:

1. Происхождение современного понимания многообразия оснований как метафизической позиции в рамках современной философии физики восходит (как минимум по духу, если не по букве) к некоторым работам позднего Э. Гуссерля [15]. Оправдано выделение таких моментов аргументации заявленного тезиса:

1.1. Поздние работы Э. Гуссерля направлены на критику оснований естественных наук, но они пересматривают и ранние эпистемологические представления самого Э. Гуссерля [16]. Сходная позиция Л. Витгенштейна [17], в особенности концепция языковой игры, стала широко известна уже после поздних работ Э. Гуссерля, хотя затруднительно в этом увидеть обязательное свидетельство преемственности. Тем не менее в «Логических исследованиях» Э. Гуссерль полагает, что достижение достоверности знания не зависит от влияния окружающего мира. Чистая логика может раскрывать такую достоверность. Логика обеспечивает единство основ в научном познании. В более позднем «Кризисе европейского человечества» Э. Гуссерль предлагает включить в понятие научной рациональности некоторые нередуцируемые (интуитивно данные) аспекты жизненного мира ученого [15. P. 183]. Таким образом, на раннем этапе Э. Гуссерль мало обращал внимания на время как основу исследовательской деятельности (логические истины связываются безотносительно контекстов их употребления). Идея жизненного мира на позднем этапе без времени немыслима.

1.2. Разработка комплекса идей, в центре которого концепт жизненного мира, приводит Э. Гуссерля к положению, что так называемые духовные болезни, которые влекут за собой войны и взаимное непонимание между народами, предполагают напоминание не только о значении интуитивного жизненного мира в науке, но и указание на поиск релевантных ему методов познания в историческом времени. Современная наука и философия не могут решить эту проблему, потому что наука и философия подменяют действительный мир искусственными структурами, такими как математические уравнения, а философия – узко понятым человеческим разумом. Возможность раскола открывается как трансцендентальная характеристика научного познания. Многообразие оснований рождается под влиянием образа знания как продукта жизненного мира, которым каждый ученый интуитивно руководствуется в актах сознания.

1.3. Необходимость исторического поворота в обсуждении проблемы дает автору возможность оспорить, например, идеи Т. Ненона (Nenon), достаточно аисторично полагающего, что трансцендентализм в целом, которого придерживается Э. Гуссерль, и классические трансценденталистские концепции, одну из которых активно развивал И. Кант, фундаментально не отличаются [18. P. 431]. Автор данной статьи считает, что трансцендентализм И. Канта и Э. Гуссерля все-таки фундаментально отличается. Э. Гуссерль выходит за рамки попыток кантовской объективации времени и пространства, т.е. хоть и фиксируется интуиция, сходная с той, что есть и в трансцендентальной философии И. Канта, но демонстрируется не одна лишь иная форма аргументации. Э. Гуссерль ищет иной основы знаний, нежели И. Кант, потому-то и ответ на вопрос относительно оснований познавательной деятельности у Э. Гуссерля оригинален. С этой точки зрения время (и пространство) не может иметь «объективно субъективного» характера, принадлежа субъекту неотчуждаемо и тем самым оказываясь одним из атрибутов познающего субъекта, который безотносителен другим его характеристикам, таким как свобода, например. Субъективность времени именно что субъективна (относительна), а не субъектна, т.е. безотносительна к другим проявлениям субъекта как такового. Дело, видимо, в том, что мысль И. Канта двигалась в за-фиксированной И. Ньютоном перспективе абсолютистских представлений о пространстве и времени, в то время как Э. Гуссерль мыслил, так сказать, в релятивистской Вселенной А. Эйнштейна.

2. Релятивистские отсылки идей Э. Гуссерля (в смысле их связи с теорией относительности в ее физическом и философском смыслах) влияют на способы понимания науки и ученого, даже если отдельные исследователи не согласны с прочими воззрениями Э. Гуссерля или даже если его воззрения не известны таким исследователям во всех деталях. На это указывают, например, любопытные замечания некоторых современных авторов, специализирующихся, правда, в области исторически ориентированной социологии науки [19. P. 236]. Проблема времени в этом контексте отсылает к поиску основ человеческой действительности [20], и автор считает, что такое значение проблемы времени помогает найти образ единства для современной эпистемологии как знания об основаниях научного познания. Другое дело, что концепт «жизненного мира» отсылает к «относительным» часам, а обоснование через него научной деятельности требует интерпретации тех же часов в универсалистском духе (научных часов много, но суть многообразия едина для всех таких часов).

Выявленный символизм Универсальных Часов позволяет приступить к обсуждению следующего тезиса:

*Т.4.* Символизм Универсальных Часов раскрывает сущность отношений, с одной стороны, между философскими спекуляциями, а также естествознанием и гуманитарным знанием – с другой.

Предлагаются следующие доводы в пользу выдвинутого тезиса:

Важно уточнить, что аргументация, направленная на подтверждение сформулированного выше тезиса *Т.0,* предполагает, что проблема времени, которую автор статьи раскрывает в виде поиска Универсальных Часов, позволяет обратить особое внимание на отношения между философскими умозрениями и исследованиями в сфере естествознания в целом и физики в частности. Намеченная связь символизма Универсальных Часов с интерпретацией времени в естествознании и гуманитарных дисциплинах, преимущественно философии, позволяет обратить внимание на то, что проблема времени влечет за собой вопрос об условиях, при которых достигается тождество естественного мира и человеческого бытия, т.е. согласование сфер объективного и субъективного служит базисом научных знаний. Философы XX столетия заключают, что достичь рационального решения вопроса о конкретном основании науки довольно сложно, будь то феноменология сознания или сохранение традиционного противоречия между сознанием и действительностью. Решение задачи может быть осуществленено как итог прагматичного выбора, выполненного на базе произвольного метафизического постулата [21. P. 625]. Неклассический подход здесь подразумевает представление о том, что окружающий мир и его атрибуты являются лишь эпифеноменом человеческого сознания, и эти идеи демонстрируют интересные параллели с теорией относительности А. Эйнштейна, которая основывается на понятии пространственно-временного континуума Г. Минковского [22].

Символизация времени в виде Универсальных Часов позволяет внести ясность в единство оснований философии и физики, ориентируясь на нормы философии науки в целом, которая охватывает как проблемы естествознания, так и вопросы гуманитарного знания. В целом время не исчерпывается по смыслу ни одним, ни другим пониманием, но выступает в качестве одной из искусственных мер, применяемых человеком. В том же отношении достигается и решение еще одной проблемы: современное расхождение между основаниями отдельных научных дисциплин существенно сокращается, если не исчезает вовсе.

## Заключение

Символическое значение понятия времени демонстрирует предпосылки для организации оснований как естественных наук, так и гуманитарных знаний в двух отношениях:

1. Связь формирования образа человеческой природы в современной философии с физическими изысканиями открывает эпистемологическую роль понятия времени, возможность использовать время как базис обобщенной теории познании.

2. Проведенное исследование раскрывает перспективу осмысления естественных и культурных процессов, основанную на объединении относительно независимых комплексов естественных наук, а также гуманитарных знаний, их активном взаимодействии и взаимном обогащении в рамках наработок современной философии физики в сфере построения моделей понимания времени.

Вполне возможно, что при таком подходе не удается решить сугубо метафизическую и потому «вечную» проблему философии «что есть время?», но становится возможным обсудить целый ряд методологических нюансов, связанных с проблемой времени в философии, физико-математическом естествознании и гуманитарных дисциплинах. Вместе с тем, как замечает В.А. Суровцев, «аналитическая философия показала, что внимание к нюансу в философии не менее важно, чем стремление к всеобщему, и именно внимание к нюансу может быть наиболее плодотворно» [23]. Поэтому автор надеется, что достигнутые результаты, даже если и вызовут значительную критику со стороны представителей других направлений философии, все-таки не будут проигнорированы и отвергнуты *a priori*.

### *Литература*

***Sergey B. Kulikov,*** Tomsk State Pedagogical University (Tomsk, Russian Federation).
E-mail: kulikovsb@tspu.edu.ru

**PHILOSOPHY ON TIME IN THE NATURAL SCIENCES AND THE HUMANITIES FROM ANALYTICAL POSITIONS**

This article aims at applying the approaches peculiar to analytic philosophy to the question about representation of the concept of time as a symbol which can reflect the bases of the modern natural sciences, social sciences and humanities. The main methods, which the author of this article uses, are speculative analysis and modeling. The symbolic meaning of the concept of time demonstrates preconditions for the organization of the bases of the natural sciences, and social and humanitarian knowledge as well. Judgments for the meaning of time reveal the essence of the problem in two aspects of discussion on the dissociation of the foundations in the modern philosophy of physics and the philosophical analysis of the humanities as well. 1) The formation of the image of human nature in contemporary philosophy reveals the special role of the concept of time in epistemology and philosophy of science. 2) This research reveals the perspective of understanding natural and cultural processes, which is based on the unification of branches of science. As a result, the research shows the basics of communication of the natural sciences with the social science, and humanitarian knowledge as well. This way, the problem of whether time is a natural process or it is only a human invention can be solved. In general, time is not exhausted by the meaning of either of the understandings, and acts as one of the artificial measures people apply. In the same way, the solution of another problem is reached: the modern discrepancy between separate bases of science decreases considerably, if not disappears completely.